\RequirePackage{fix-cm}
\documentclass[smallextended]{svjour3}       
\smartqed  
\usepackage{graphicx}
\begin{document}

\title{Thermodynamics of Asymptotically Flat Black Holes in Lovelock Background
}


\author{N. Abbasvandi         \and
        M. J. Soleimani  \and Shahidan Radiman \and W.A.T. Wan Abdullah \and G. Gopir
}


\institute{ N. Abbasvandi \at
        	School of Applied Physics, FST, University Kebangsaan Malaysia, Bangi, 43600, Malaysia
        	Tel.: +603-8921-5419
        	Fax: +603-8925-6086
        	\email{Niloofar@siswa.ukm.edu.my}
         \and
        	M. J. Soleimani \at
              Physics Department, University of Malaya, KL, 50603, Malaysia \\
              Tel.: +603-7967 4201\\
              Fax: +603-7956 6343\\
              \email{msoleima@cern.ch}           
           \and
           Shahidan Radiman \at
           School of Applied Physics, FST, University Kebangsaan Malaysia, Bangi, 43600, Malaysia
           Tel.: +603-8921-5419
           Fax: +603-8925-6086
           \email{Shahidan@ukm.edu.my}
           \and
           W.A.T. Wan Abdullah \at
           Physics Department, University of Malaya, KL, 50603, Malaysia \\
           Tel.: +603-7967 4201\\
           Fax: +603-7956 6343\\
           \email{wat@um.edu.my}
           \and
           G. Gopir \at
           School of Applied Physics, FST, University Kebangsaan Malaysia, Bangi, 43600, Malaysia
           Tel.: +603-8921-5419
           Fax: +603-8925-6086
           \email{gkagopir@ukm.edu.my}
}

\date{Received: date / Accepted: date}

\maketitle

\begin{abstract}
We study the thermodynamics of the asymptotically flat static black hole in Lovelock back ground where the coupling constants of the Lovelock theory effects are taken into account. We consider the effects of the second order of the coupling constant, and third order of the Lovelock constant coefficient on the thermodynamics of asymptotically flat static black holes. In this case the effect of the coupling constants on the thermodynamics of the black hole are discussed for 5, 6, and 7 dimensional spacetime.
\keywords{Lovelock Theory \and Thermodynamics \and Hawking temperature \and Entropy \and Higher Dimensions}
\PACS{04.50.Kd, \and  04.70.Dy}
\end{abstract}

\section{Introduction}
\label{sec:intro}
One of the possible solutions to the hierarchy problem, is a class of models in which extra spatial dimensions \cite{arkani,antoniadis,arkanni2,randall} give rise to strong gravity due to the fact that in $4 + n$ dimensions the fundamental Planck mass (MD) can be as low as few TeV. The most significant consequence of the low-scale gravity, is the possibility of production of the TeV-scale black holes in particle colliers, such as Large Hadron Collier (LHC) \cite{greg} and the available energy must be well above the fundamental Planck scale \cite{gidding}. In string theory, in addition to Einstein-Hilbert term, there are higher curvature corrections \cite{boulware}. On the other hand, higher dimensional black holes  have been used to analyzed strongly coupled finite temperature field theories. So, extension of Einstein gravitational theory into those with higher power of curvature in higher dimensions is natural. In this regards, the Lovelock theory of gravity is a candidate belonging to such class of theories \cite{lovelock,charmousis} which had received a lot of attention, in particular, in the brane world scenario, low energy string theory, and black hole thermodynamics. Since the thermodynamic property of black hole is essentially a quantum feature of gravity, it is expected to gain some insights into quantum gravity in the field of black hole thermodynamics. 
In this case, the effective action proposed by Lovelock \cite{lovelock} contains the higher powers of curvature in particular combinations and the only produced field equations are the second order one which consequently arise no ghosts \cite{zwiebach}. In this case, the action is precisely of the form
\begin{equation} \label{1}
{I_G} = \int {{d^d}x\sqrt { - g} L} 
\end{equation}
where $L$ denotes the corresponding Lagrangian.
\\Here, we consider the first four terms of the Lovelock gravity which is up to the third order. The combination of these terms in $d \ge 7$ is the most general Lagrangian producing second order field equations. It is very difficult to find out nontrivial exact analytical solutions of Einstein's equation with the higher curvature terms due to the non linearity of the field equations and mostly need to adopt some approximation methods or find numerical solutions. Wheeler has found \cite{wheeler} exact static spherically symmetric black hole solutions of the Gauss-Bonnet (the second order Lovelock) gravity. Also, one can find related solution of the Einstein-Maxwell-Gauss-Bonnet and Einstein-Born-Infeld-Gauss-Bonnet models in Refs \cite{wiltshire,wiltshire1}. The thermodynamics of the charged static spherically black hole solutions has been considered
in \cite{lidsey} and of charged solutions in \cite{myers}. In this paper we want to find new static solutions of third order Lovelock gravity by using algebraic equation (see Eq.(9)) which are asymptotically flat and contain second and third order coupling constant of Lovelock gravitational theory and investigate their thermodynamics. In this case, we will investigate the effect of the coupling constants on Hawking temperature and Bekenstein-Hawking entropy as well, specially in 7 and 8 dimensional spacetime.
\\The outline of the paper is as follows. In section 2, we briefly review Lovelock theory and we obtain general algebraic equation. In section 3, we present the static solution of the third order Lovelock theory and investigate the thermodynamic parameters for seven dimensional spacetime. We obtain the solution for eight dimensional spacetime and investigate its thermodynamical properties in section 4. The final section 5 is devoted to conclusion and discussion. 

\section{Lovelock Black Hole Spacetime}
One of the natural generalization of Einstein theory in higher dimensional spacetime is Lovelock gravity \cite{lovelock} which maintains the properties of Einstein equation. Thus, for Eq. (1), the effective action, we consider third order Lovelock black hole with a cosmological constant term, where the corresponding Lagrangian is given by
\begin{equation} \label{2}
L = \sum\limits_{m = 0}^{{{\left( {D - 1} \right)} \mathord{\left/
			{\vphantom {{\left( {D - 1} \right)} 2}} \right.
			\kern-\nulldelimiterspace} 2}} {{c_m}{\ell _m}}
\end{equation}
where ${{c_m}}$ is an arbitrary constant and ${{\ell _m}}$ is the Euler density of $2k$-dimensional manifolds which is defined by
\begin{equation} \label{3}
{\ell _m} = \frac{1}{{{2^m}}}\delta _{{\mu _1}{\nu _1}...{\mu _m}{\nu _m}}^{{\lambda _1}{\xi _1}...{\lambda _m}{\xi _m}}{R_{{\lambda _1}{\xi _1}}}^{{\mu _1}{\nu _1}}...{R_{{\lambda _m}{\xi _m}}}^{{\mu _m}{\nu _m}}.
\end{equation}
In Eq. (3), $\delta _{{\mu _1}{\nu _1}...{\mu _m}{\nu _m}}^{{\lambda _1}{\xi _1}...{\lambda _m}{\xi _m}}$ and ${R_{{\lambda _1}{\xi _1}}}^{{\mu _1}{\nu _1}}$ are the generalized totally antisymmetric Kronecker delta and the Reimann tensor in $D$ dimension respectively. We would like to draw attention that in $d$ dimensions, all terms for which $k \succ \left( {\frac{{d - 1}}{2}} \right)$ are identically equal to zero, and the term $k = \left( {\frac{{d - 1}}{2}} \right)$ is a topological term. So, the only term for which $k \prec \left( {\frac{{d - 1}}{2}} \right)$ are contributing to the field equations. Hereafter, we set ${c_0} =  - 2\Lambda$, ${c_1} = 1$, ${c_2} = {\alpha  \mathord{\left/{\vphantom {\alpha  2}} \right.\kern-\nulldelimiterspace} 2}$, ${c_3} = {\beta  \mathord{\left/{\vphantom {\beta 3}}\right.\kern-\nulldelimiterspace} 3}$, ${c_m} = 0$ (for $4 \le m$), and $G = \hbar  = c = 1$. So, the second and third order of the Lagrangian terms are given by 
\begin{equation} \label{2.3}
{\ell _2} = {R_{\mu \nu \rho \sigma }}{R^{\mu \nu \rho \sigma }} - 4{R_{\mu \nu }}{R^{\mu \nu }} + {R^2}
\end{equation}
and
\begin{equation} \label{2.4}
\begin{array}{l}
{\ell _3} = 24{R_{\mu \nu }}^{\rho \sigma }{R_\rho }^\nu {R_\sigma }^\nu  - 24{R_{\mu \nu }}^{\rho \sigma }{R_{\rho \eta }}^{\mu \nu }{R_\sigma }^\eta  + 3R{R_{\mu \nu }}^{\rho \sigma }{R_{\rho \sigma }}^{\mu \nu }\\
+ 8{R_{\mu \nu }}^{\rho \sigma }{R_{\rho \eta }}^{\mu \kappa }{R_{\kappa \sigma }}^{\nu \eta } + 2{R_{\mu \nu }}^{\rho \sigma }{R_{\rho \sigma }}^{\eta \kappa }{R_{\eta \kappa }}^{\mu \nu } + 16{R_\mu }^\nu {R_\nu }^\rho {R_\rho }^\mu \\
- 12R{R_\mu }^\nu {R_\nu }^\mu  + {R^3}.
\end{array}
\end{equation} 
One can then derive the Lovelock equation with respect to the metric as follows
\begin{equation} \label{2.5}
0 = {\vartheta _\lambda }^\xi  = \Lambda {\delta _\mu }^\xi  + G_\lambda ^{\left( 1 \right)\xi } + \alpha G_\lambda ^{\left( 2 \right)\xi } + \beta G_\lambda ^{\left( 3 \right)\xi }
\end{equation}
where $G_\lambda ^{\left( 1 \right)\xi }$ is the Einstein tensor and $G_\lambda ^{\left( 2 \right)\xi }$ $\&$ $G_\lambda ^{\left( 3 \right)\xi }$ are the second and the third order of Lovelock tensors respectively.   
\\It was shown \cite{wheeler,cai} that there exist static exact black hole solutions of the equation (6). In this case, the line element takes the following form
\begin{equation} \label{2.6}
d{s^2} =  - f\left( r \right)d{t^2} + {f^{ - 1}}\left( r \right)d{r^2} + {r^2}{{\bar \gamma }_{ij}}d{x^i}d{x^j}
\end{equation}
where ${{\bar \gamma }_{ij}}$ is the metric of $D-2$ dimensional constant curvature space. In order to determine the function $f\left( r \right)$ by Eq. (6), it is convenient to define a new variable $\psi \left( r \right)$ as
\begin{equation} \label{2.7}
f\left( r \right) = \zeta  - {r^2}\psi \left( r \right)
\end{equation}
with the constant curvature $\zeta  = 1,0$ or $-1$. Then, after calculating Riemann tensor components and substituting into (6), one can obtain an algebraic equation as follows
\begin{equation} \label{2.8}
\sum\limits_{m = 2}^k {\left[ {\frac{{{a_m}}}{m}\left\{ {\prod\limits_{q = 1}^{2m - 2} {\left( {n - q} \right)} } \right\}{\psi ^m}} \right]}  + \psi  - \frac{{2\Lambda }}{{n\left( {n + 1} \right)}} = \frac{\mu }{{{r^{n + 1}}}}
\end{equation}
where $\Lambda$ is cosmological constant, ${\frac{{{a_m}}}{m} = {c_m}}$, and $\mu$ is an integration constant related to $ADM$ mass as $M = \frac{{2\mu {\pi ^{{{\left( {n + 1} \right)} \mathord{\left/ {\vphantom {{\left( {n + 1} \right)} 2}} \right.
						\kern-\nulldelimiterspace} 2}}}}}{{\Gamma \left( {{{\left( {n + 1} \right)} \mathord{\left/ {\vphantom {{\left( {n + 1} \right)} 2}} \right.
					\kern-\nulldelimiterspace} 2}} \right)}}$ \cite{myers}, where we used a unit $16\pi G = 1$. In the following section we obtain the exact solution for static black holes and its thermodynamical properties for 7 dimensional spacetime. 

\section{Thermodynamics of the Lovelock Black Holes}
The most general second order differential equation which presents the solutions of gravity is the gravitational field equation of third order Lovelock gravity in seven dimensions. In order to obtain seven dimensional static asymptotically solution of third order of the Lovelock gravity, we set $\zeta = 1$ and $\Lambda = 0$ and we consider only positive mass black holes ${\mu  \succ 0}$. 
\\In the case of $n = 3$, the theory is reduced to Einstein-Gauss-Bonnet theory and its  related $f\left( r \right)$ reads
\begin{equation} \label{3.1}
f\left( r \right)=1 - \frac{{ - {r^2} + \sqrt {{r^4} + 4{\mkern 1mu} \alpha {\mkern 1mu} \mu } }}{{2\alpha }}.
\end{equation} 
So, for black hole solution (7) considering (10), one can find the Hawking temperature as the inverse of the special value of the period of the Euclidean time by using ${T_H} = \frac{{f'\left( r \right)}}{{4\pi }}\left| {_{r = {r_ + }}} \right.$ as follow
\begin{equation} \label{3.2}
T_H^5\left( {{r_ + }} \right) = \frac{1}{{4\pi }}\frac{{{r_ + }\left( {\sqrt {{r_ + }^4 + 4{\mkern 1mu} \alpha {\mkern 1mu} \mu }  - {r_ + }^2} \right)}}{{\alpha \sqrt {{r_ + }^4 + 4{\mkern 1mu} \alpha {\mkern 1mu} \mu } }}
\end{equation}
Therefore, the behavior of the Hawking temperature crucially depends on the second and third order coupling constants of the Lovelock gravity theory. The Hawking temperature always increases monotonically from $T = 0$ at $r_+ = 0$ for small horizon radius (see Figure 1).
\begin{figure}[tpb]
	\centering \begin{center} \end{center} 
	\includegraphics[width=1\textwidth,trim=-200 400 0 100,clip]{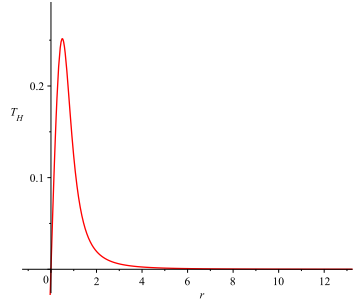}
	\hfill
	\caption{\label{fig1} Hawking temperature vs. r}
\end{figure}   
Figure 2 shows the effect of the $\alpha$ variations on the temperature. By the way, maximum of the temperature decreases by increasing of the second order coupling constant of the Lovelocke gravity in 5 dimensional asymptotically flat spacetime.
\begin{figure}[tpb]
	\centering \begin{center} \end{center} 
	\includegraphics[width=1\textwidth,trim=-200 400 0 100,clip]{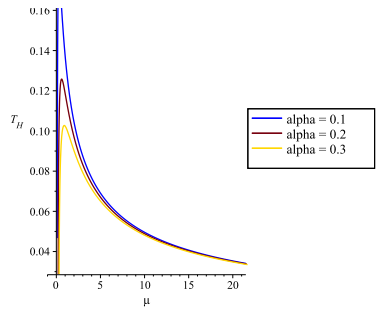}
	\hfill
	\caption{\label{fig2} Hawking temperature with respect to the different $\alpha$s for $n = 3$}
\end{figure}
we see that, as expected, (11) behaves like the Hawking temperature of a GR solution if the black hole is large enough. On the other hand, temperature tends to zero for small values.
It is easy to verify that the Hawking temperature associated to the solution in $n = 4$ ($d = 6$ dimension) is given by 
\begin{equation} \label{3.3}
T_H^6\left( {{r_ + }} \right) = \frac{1}{{4\pi }}{\mkern 1mu} \frac{{{r_ + }^2\sqrt {{r_ + }\left( {{r_ + }^5 + 12{\mkern 1mu} \alpha {\mkern 1mu} \mu } \right)}  - {r_ + }^5 + 3{\mkern 1mu} \alpha {\mkern 1mu} \mu }}{{3{r_ + }\alpha {\mkern 1mu} \sqrt {{r_ + }\left( {{r_ + }^5 + 12{\mkern 1mu} \alpha {\mkern 1mu} \mu } \right)} }}
\end{equation} 
which implies that in $d = 6$ dimensional case, the Hawking temperature is related just to the second order of the Lovelock coupling constant same as the $d = 5$ dimensional case. It is worth pointing out that for dimension 6 the functional form of the
temperature is substantially different from the case $d = 6$, as it includes an additional term which is actually proportional to $\alpha$. In this case, figure 3 shows that the temperature decreases by increasing the $\alpha$.  
\begin{figure}[tpb]
	\centering \begin{center} \end{center} 
	\includegraphics[width=1\textwidth,trim=-200 400 0 100,clip]{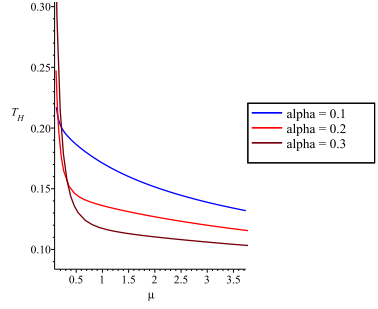}
	\hfill
	\caption{\label{fig2} Hawking temperature with respect to the different $\alpha$s for $n = 3$}
\end{figure}
In order to obtain the temperature for $d = 7$ dimension, substituting $n = 5$ into Eq. (9), we obtain a simple equation as
\begin{equation} \label{3.4}
8{\mkern 1mu} \beta {\mkern 1mu} {\Psi ^3} + 6{\mkern 1mu} \alpha {\mkern 1mu} {\Psi ^2} + \Psi  = \frac{\mu }{{{r^6}}}
\end{equation}
Then, one can easily find $\psi$ and by using Eq. (8) obtain the function $f\left( r \right)$ as follows
\begin{equation} \label{3.5}
\begin{array}{l}
f\left( r \right) = 1 - {\mkern 1mu} \frac{{{{\left( { - 27{\mkern 1mu} {\alpha ^3}{r^6} + 27{\mkern 1mu} \alpha {\mkern 1mu} {r^6}\beta  + 108{\mkern 1mu} \mu {\mkern 1mu} {\beta ^2} + 3{\mkern 1mu} \xi \beta } \right)}^{\frac{1}{3}}}}}{{12\beta }} - {\mkern 1mu} \frac{{{r^4}\left( {3{\mkern 1mu} {\alpha ^2} - 2{\mkern 1mu} \beta } \right)}}{{4\beta {\mkern 1mu} {{\left( { - 27{\mkern 1mu} {\alpha ^3}{r^6} + 27{\mkern 1mu} \alpha {\mkern 1mu} {r^6}\beta  + 108{\mkern 1mu} \mu {\mkern 1mu} {\beta ^2} + 3{\mkern 1mu} \xi \beta } \right)}^{\frac{1}{3}}}}}\\
\begin{array}{*{20}{c}}
{\begin{array}{*{20}{c}}
	{}&{}&{}
	\end{array}}&{}&{}
\end{array} + {\mkern 1mu} \frac{{\alpha {\mkern 1mu} {r^2}}}{{4\beta }}
\end{array}
\end{equation} 
where $\xi  = {\left( { - 27{\mkern 1mu} {\alpha ^2}{r^{12}} + 24{\mkern 1mu} \beta {\mkern 1mu} {r^{12}} - 648{\mkern 1mu} {\alpha ^3}\mu {\mkern 1mu} {r^6} + 648{\mkern 1mu} \alpha {\mkern 1mu} \beta {\mkern 1mu} \mu {\mkern 1mu} {r^6} + 1296{\mkern 1mu} {\beta ^2}{\mu ^2}} \right)^{\frac{1}{2}}}$ and solution (10) has only one branch unlike the solutions in Gauss-Bonnet gravity which have two branches. Indeed, Eq. (10) with above $\alpha$s and $\beta$s has one real solution and two complex solutions. 
\\The Hawking temperature of the black holes can be obtained by requiring the absence of conical singularity at the event horizon in the Euclidean sector of the black hole solution which can be obtained by analytical continuation of the metric. We continue analytically the metric by $t \to i\tau$ which yields the Euclidean section. The regularity at $r = {r_ + }$ requires that we should identify $\tau  \sim \tau  + {\beta _ + }$ where $f\left( {{r_ + }} \right) = 0$ and ${\beta _ + }$ is the inverse Hawking temperature of the event horizon given as
\begin{equation} \label{3.6}
\beta _ + ^{ - 1} = T_H^7\left( {{r_ + }} \right) = \frac{1}{{4\pi }}\left( \begin{array}{l}
{\mkern 1mu} \frac{{{r_ + }^4\left( {3{\mkern 1mu} {\alpha ^2} - 2{\mkern 1mu} \beta } \right)\gamma }}{{12\beta {\mkern 1mu} {{\left( { - 27{\mkern 1mu} {\alpha ^3}{r_ + }^6 + 27{\mkern 1mu} \alpha {\mkern 1mu} {r_ + }^6\beta  + 108{\mkern 1mu} \mu {\mkern 1mu} {\beta ^2} + 3{\mkern 1mu} \xi {\mkern 1mu} \beta } \right)}^{4/3}}}}\\
- \frac{{{r_ + }^3\left( {3{\mkern 1mu} {\alpha ^2} - 2{\mkern 1mu} \beta } \right)}}{{\beta {\mkern 1mu} {{\left( { - 27{\mkern 1mu} {\alpha ^3}{r_ + }^6 + 27{\mkern 1mu} \alpha {\mkern 1mu} {r_ + }^6\beta  + 108{\mkern 1mu} \mu {\mkern 1mu} {\beta ^2} + 3{\mkern 1mu} \xi {\mkern 1mu} \beta } \right)}^{\frac{2}{3}}}}}\\
+ {\mkern 1mu} \frac{{ - \gamma }}{{36{{\left( { - 27{\mkern 1mu} {\alpha ^3}{r_ + }^6 + 27{\mkern 1mu} \alpha {\mkern 1mu} {r_ + }^6\beta  + 108{\mkern 1mu} \mu {\mkern 1mu} {\beta ^2} + 3{\mkern 1mu} \xi {\mkern 1mu} \beta } \right)}^{2/3}}\beta }}\\
+ {\mkern 1mu} \frac{{\alpha {\mkern 1mu} {r_ + }}}{{2\beta }}
\end{array} \right)
\end{equation}
where
\begin{equation} \label{3.7}
\begin{array}{l}
\gamma  =  - 162{\mkern 1mu} {\alpha ^3}{r_ + }^5 + 162{\mkern 1mu} \alpha {\mkern 1mu} {r_ + }^5\beta  + {\mkern 1mu} \frac{{3\beta {\mkern 1mu} \left( { - 324{\mkern 1mu} {\alpha ^2}{r_ + }^{11} + 288{\mkern 1mu} \beta {\mkern 1mu} {r_ + }^{11} - 3888{\mkern 1mu} {\alpha ^3}\mu {\mkern 1mu} {r_ + }^5 + 3888{\mkern 1mu} \alpha {\mkern 1mu} \beta {\mkern 1mu} \mu {\mkern 1mu} {r_ + }^5} \right)}}{{2\xi }}\\
\xi  = {\left( { - 27{\mkern 1mu} {\alpha ^2}{r_ + }^{12} + 24{\mkern 1mu} \beta {\mkern 1mu} {r_ + }^{12} - 648{\mkern 1mu} {\alpha ^3}\mu {\mkern 1mu} {r_ + }^6 + 648{\mkern 1mu} \alpha {\mkern 1mu} \beta {\mkern 1mu} \mu {\mkern 1mu} {r_ + }^6 + 1296{\mkern 1mu} {\beta ^2}{\mu ^2}} \right)^{\frac{1}{2}}}.
\end{array}
\end{equation}
Clearly here , the temperature is related to the second order coupling constant as the 5 and 6 dimensional case. Also, it depends on the third order constant of Lovelock as well as second order. Figure 4 shows the $\alpha$ effects on the Hawking temperature of fixed $\beta$. We also show the effect of $\beta$ of fixed $\alpha$. It is seen that, the Hawking temperature increases when the second order constant of the Lovelock increases but the Hawking temperature decreases when the third order constant of the Lovelock increases (see Figure 4 and 5). We have shown that in the absence of the second order constant, $\alpha$, the third order constant of the Lovelock, $\beta$, still decreases the Hawking temperature (Figure 6). 
\begin{figure}[tbp] \label{4}
	\centering 
	\includegraphics[width=.4\textwidth,origin=a,trim=-90 70 200 100,]{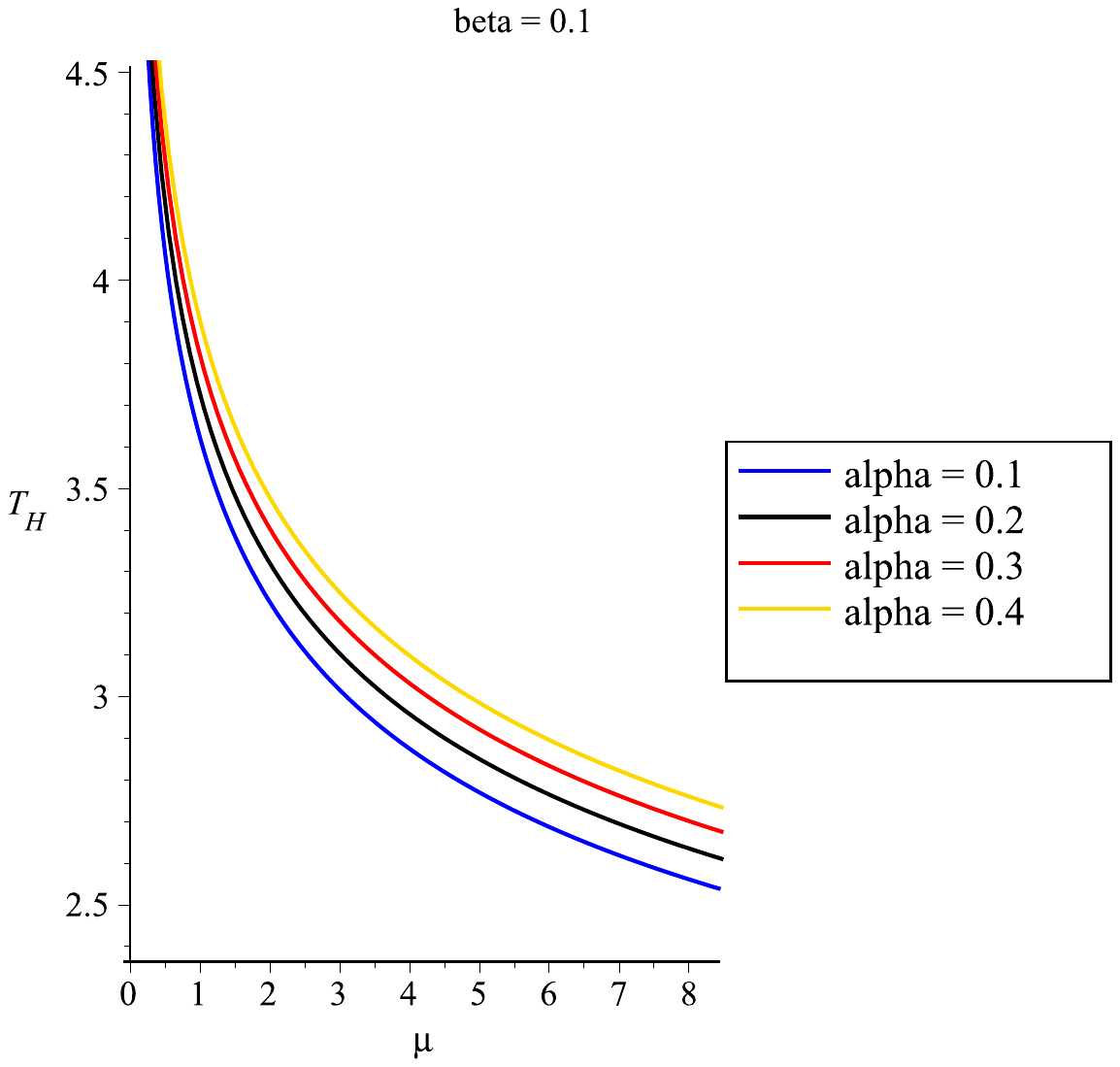}
	\hfill 
	\includegraphics[width=.4\textwidth,origin=a,trim=-90 70 200 100,]{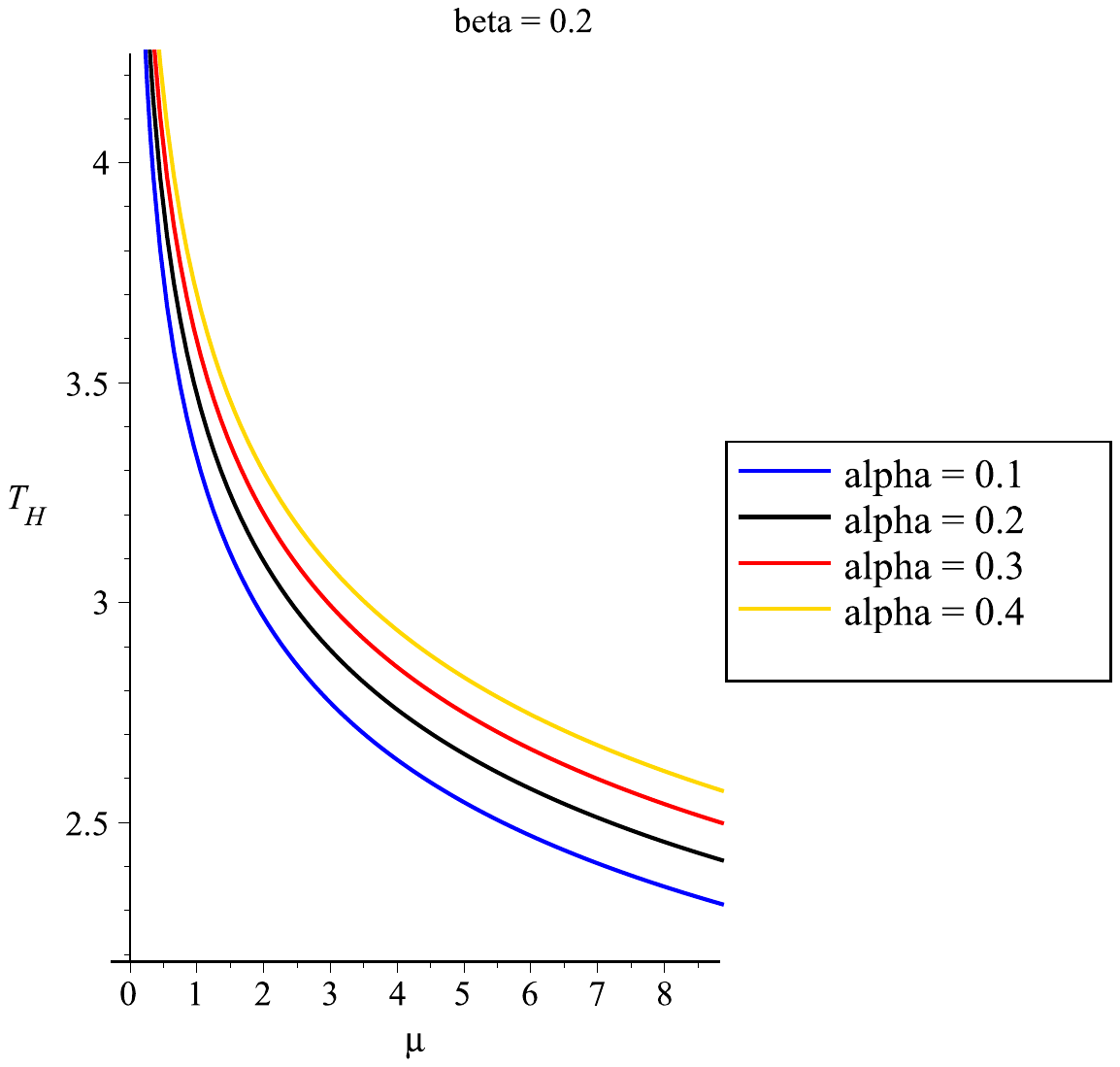}
	\hfill
	\includegraphics[width=.4\textwidth,origin=a,trim=-90 70 200 100, angle=0]{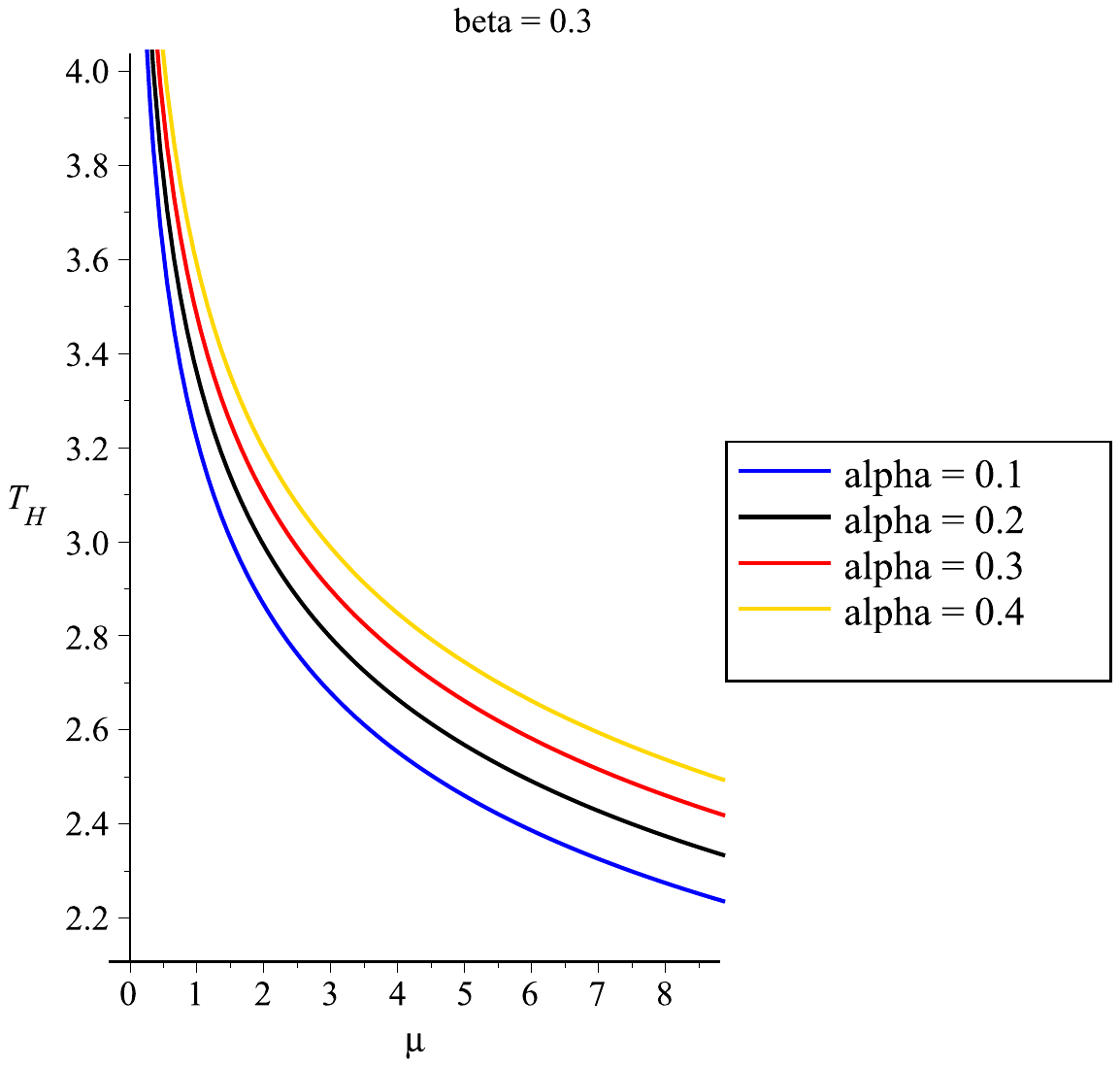}
	\hfill
	\includegraphics[width=.4\textwidth,origin=a,trim=-90 70 200 60, angle=0]{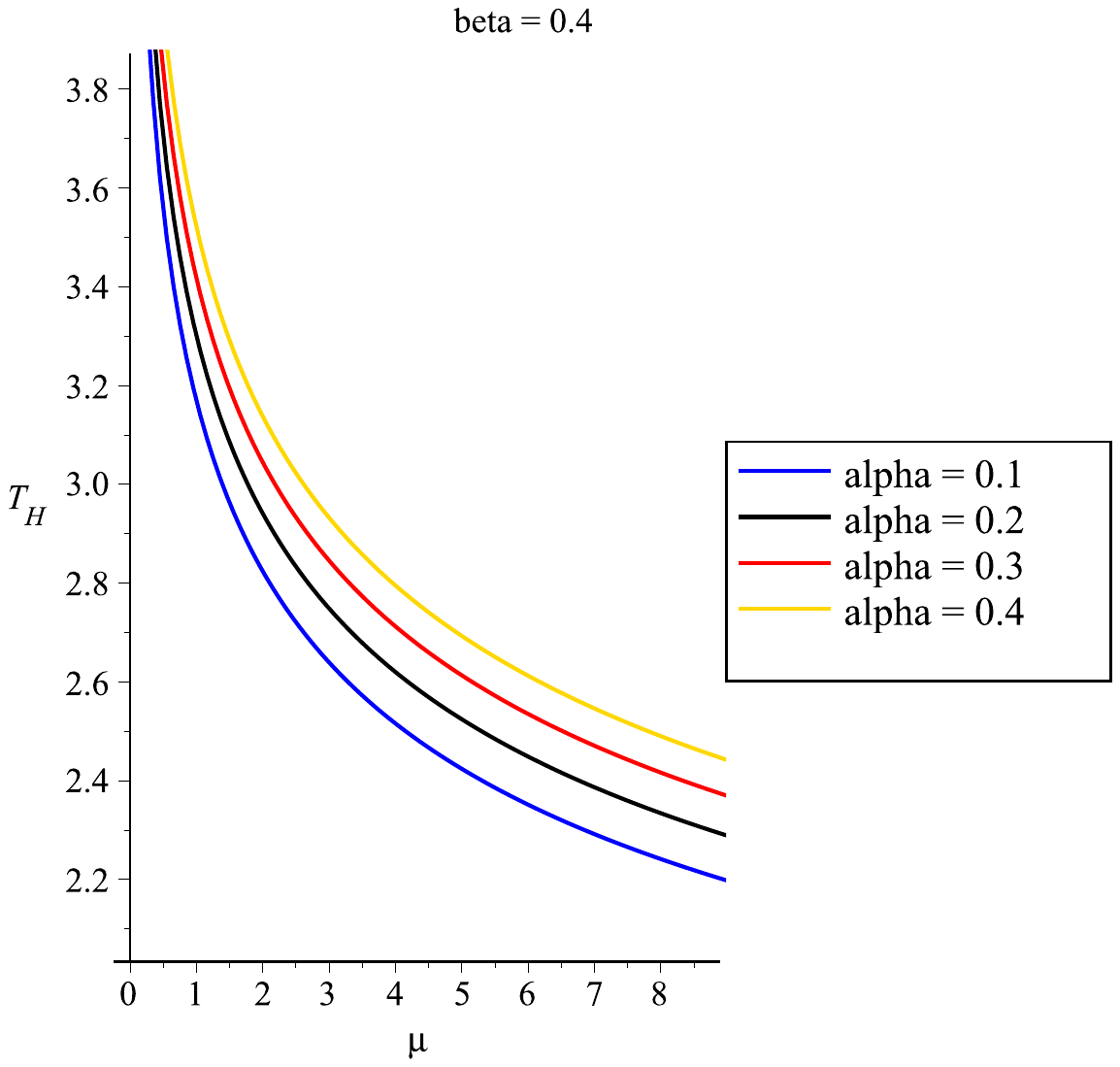}
	\hfill
	
	\caption{\label{fig:4} The $\alpha$'s effects on the Hawking temperature for different $\beta$ .}
\end{figure}
\begin{figure}[tbp] \label{5}
	\centering 
	\includegraphics[width=.4\textwidth,origin=a,trim=-90 70 200 100,]{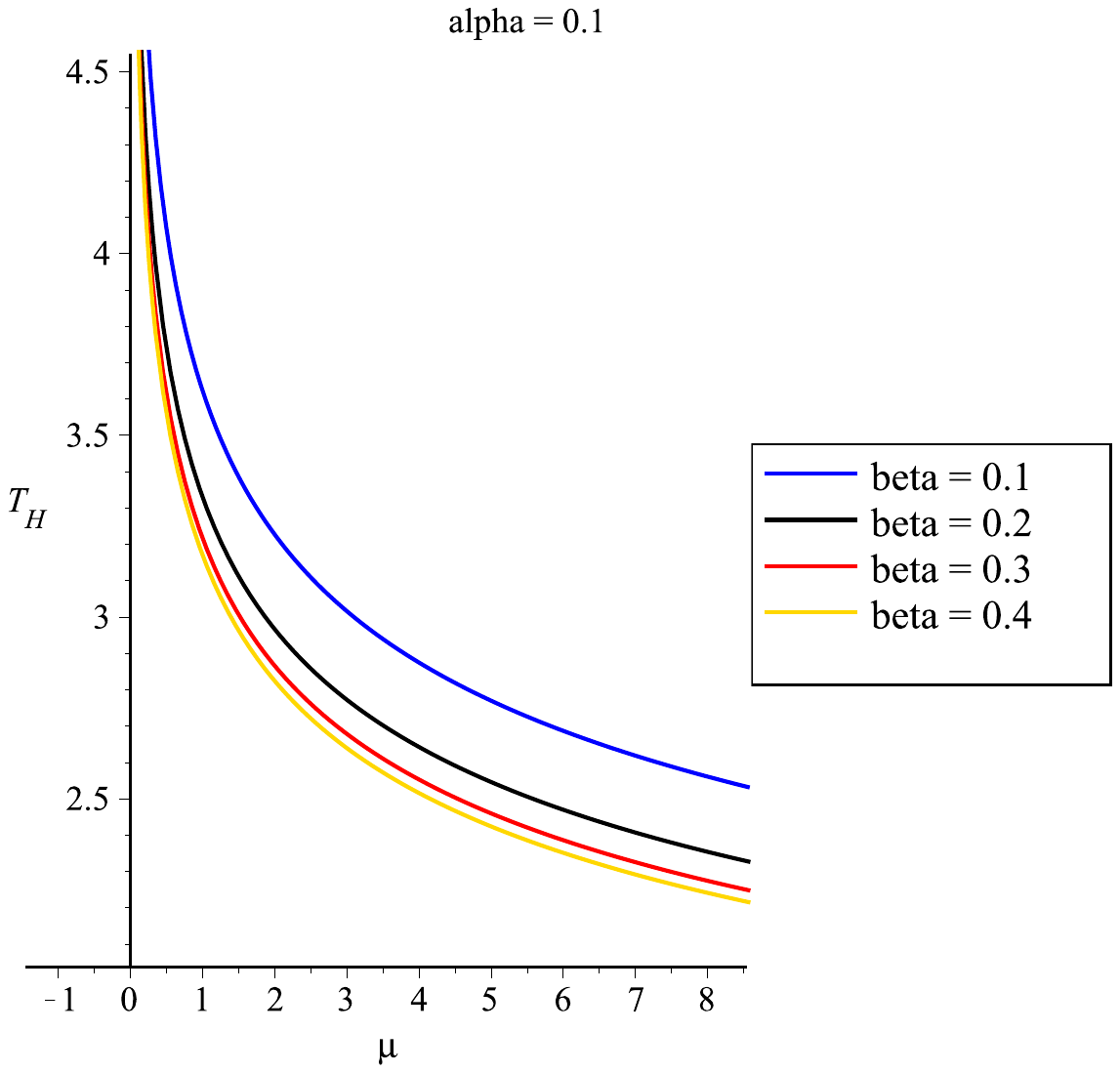}
	\hfill 
	\includegraphics[width=.4\textwidth,origin=a,trim=-90 70 200 100,]{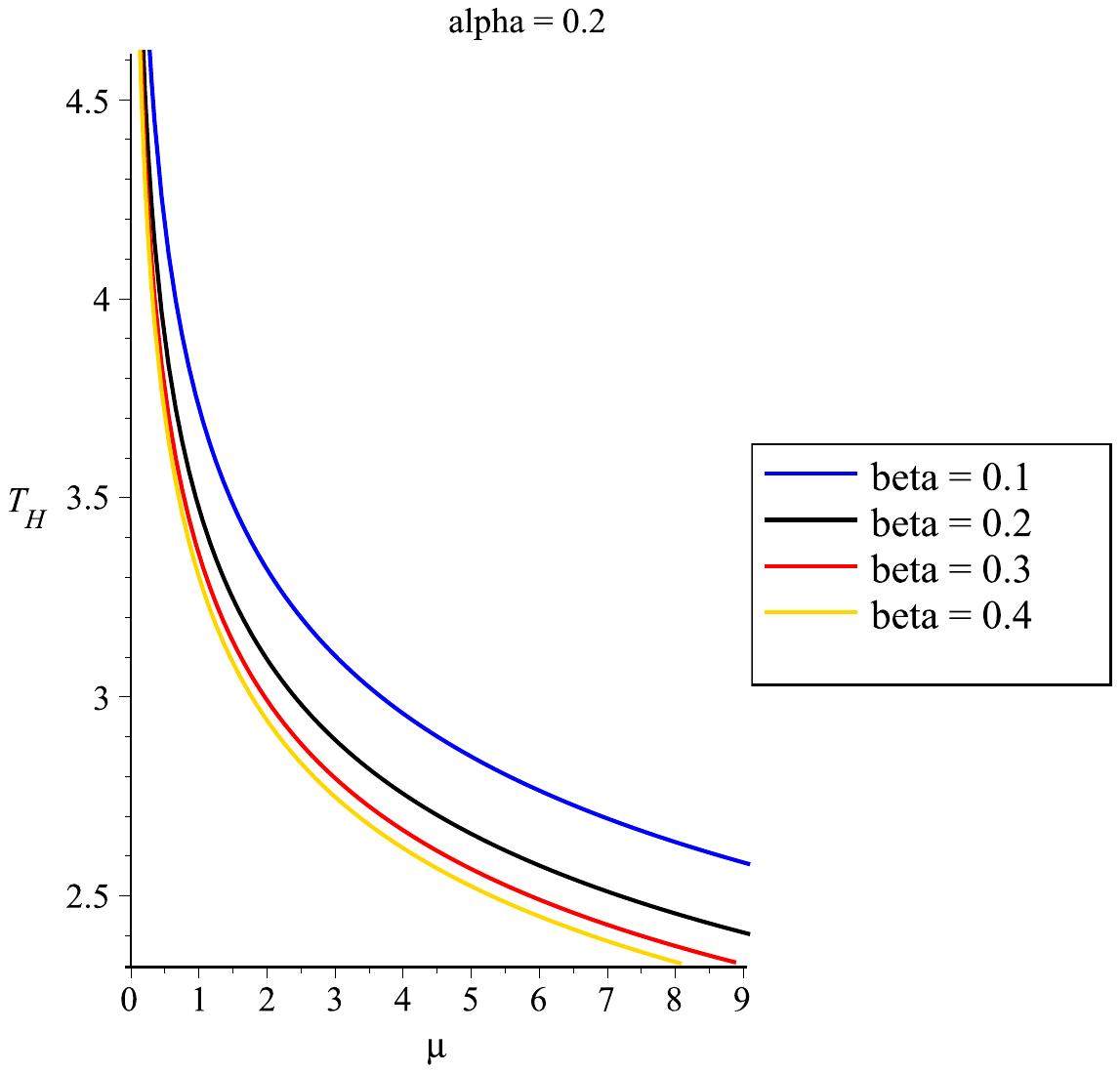}
	\hfill
	\includegraphics[width=.4\textwidth,origin=a,trim=-90 70 200 100, angle=0]{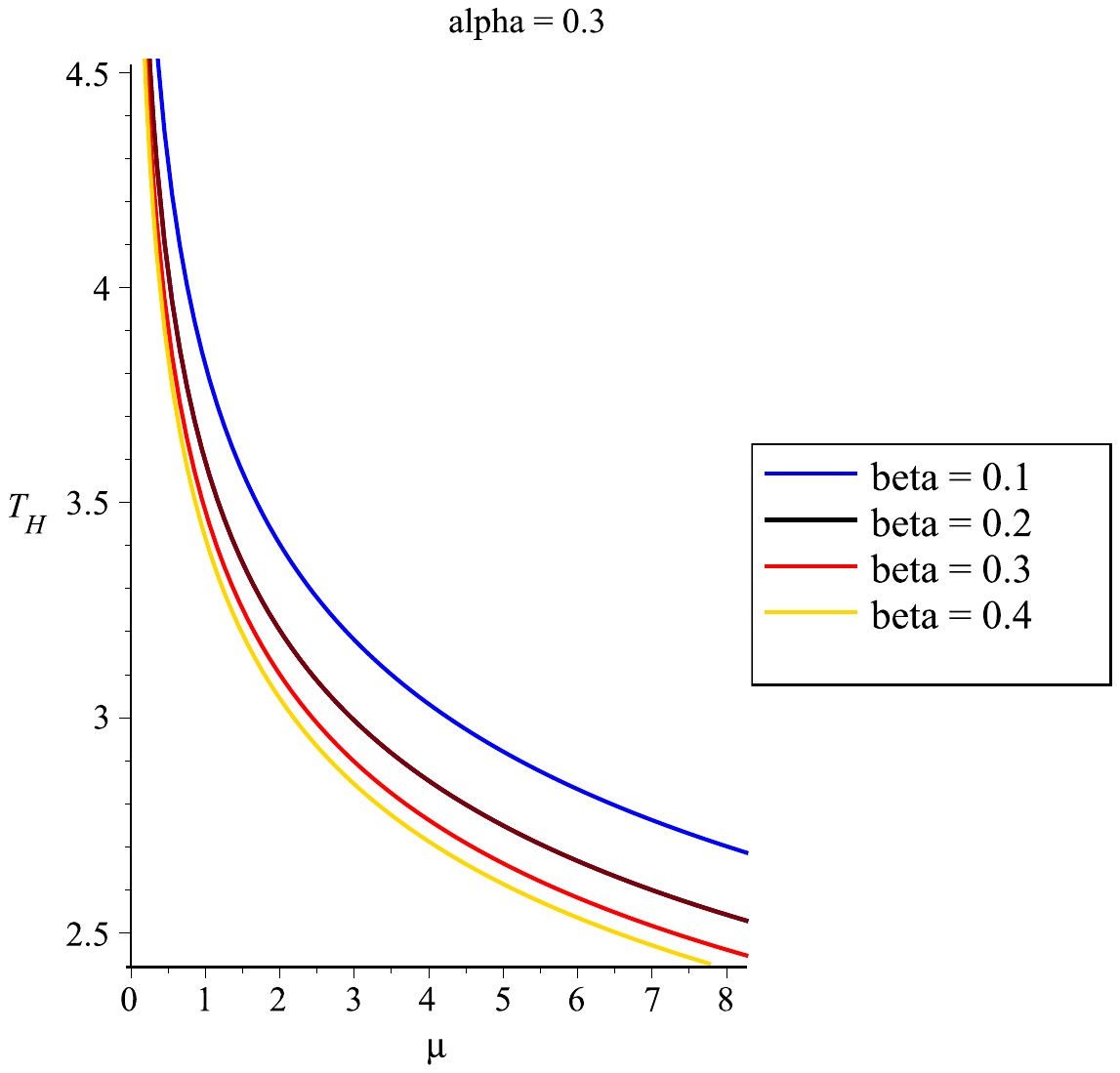}
	\hfill
	\includegraphics[width=.4\textwidth,origin=a,trim=-90 70 200 60, angle=0]{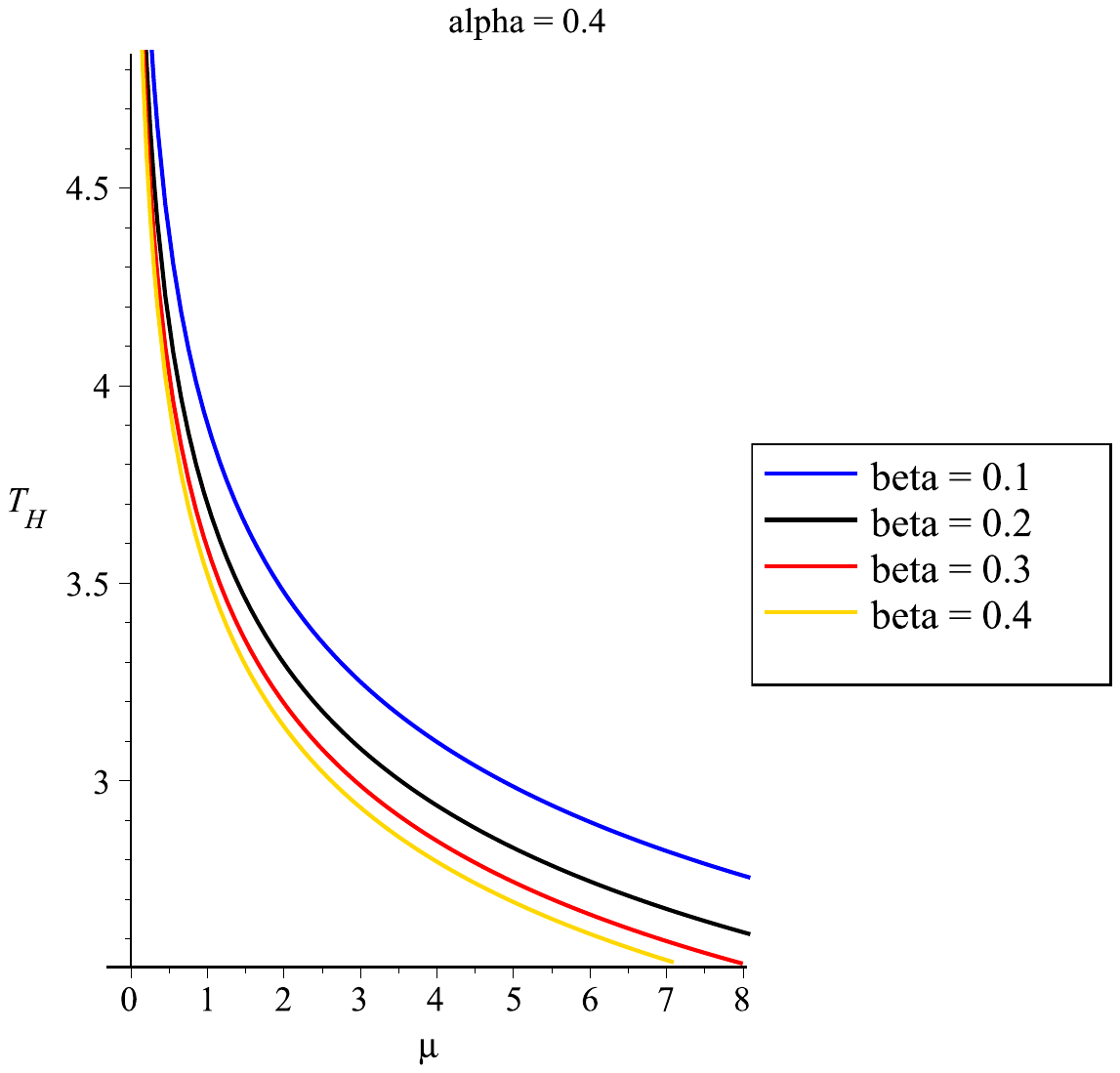}
	\hfill
	
	\caption{\label{fig:5} The $\beta$'s effects on the Hawking temperature for different $\alpha$ .}
\end{figure}
\begin{figure}[tpb]
	\centering \begin{center} \end{center} 
	\includegraphics[width=1\textwidth,trim=-200 400 0 100,clip]{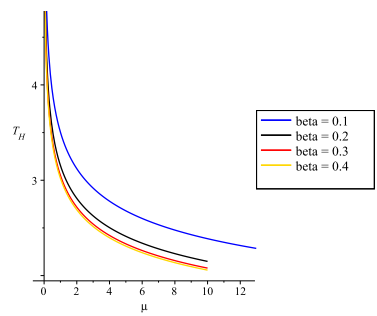}
	\hfill
	\caption{\label{fig6} The effect of $\beta$ on the Hawking temperature in the absence of the $\alpha$}
\end{figure}
\\In general, the Bekenstein-Hawking entropy of black holes satisfies the so-called area law of entropy which states that the black hole entropy equals one-quarter of the horizon area, $S = \frac{A}{4}$ \cite{bekenstein}. However, in higher derivative gravity the area law of entropy is not satisfied mostly \cite{lu}.The entropy per unit volume obeys the law of the
entropy of asymptotically flat black holes of pth-order Lovelock gravity \cite{jacobson}
\begin{equation} \label{3.8}
S = \frac{1}{4}\sum\limits_{k = 1}^p {k{\alpha _k}\int {{d^{n - 1}}x\sqrt {\tilde g} {{\tilde l}_{k - 1}}} }
\end{equation} 
where the integration is done on the $(n-1)$-dimensional space like hypersurface of Killing horizon, ${{{\tilde l}_{k - 1}}}$is the kth order Lovelock Lagrangian of ${{\tilde g}_{\mu \nu }}$ which is the induced metric on it and ${\tilde g}$ is the determinant of ${{\tilde g}_{\mu \nu }}$. By the way, a simple method of finding the entropy is through the use of first law of thermodynamics, $dM = TdS$ \cite{wald}. The horizon radius s
tarts from zero when the mass parameter is nonnegative. So, by integrating the first law, the entropy reads
\begin{equation} \label{3.9}
S = \int\limits_0^{{r_ + }} {{T^{ - 1}}\left( {\frac{{\partial M}}{{\partial {r_ + }}}} \right)d{r_ + }}.
\end{equation}
in this case, one can obtain the entropy for $n = 3$ as follows
\begin{equation} \label{3.10}
{S^5} = 8{\mkern 1mu} \frac{{\pi {\mkern 1mu} \left( {2/5{\mkern 1mu} {r_ + }^5 + \alpha {\mkern 1mu} \mu {\mkern 1mu} {r_ + }} \right)}}{\mu }
\end{equation}
which is related to the $\alpha$ as second order constant of the Lovelock theory. Figure 7 shows the entropy deviation by variation of the $\alpha$ and the entropy decreases when the $\alpha$ increases.

In the same way, one can compute the entropy for 6 dimensional asymptotically flat black hole in Lovelock back ground as follows
\begin{equation} \label{3.11}
\begin{array}{l}
{S^6} = 1/4{\mkern 1mu} \frac{{\pi {\mkern 1mu} {r_ + }^6}}{\mu } + \frac{{3{\mkern 1mu} \pi {\mkern 1mu} {r_ + }^5}}{{80{\mkern 1mu} \mu }} + \frac{{3{\mkern 1mu} \pi {\mkern 1mu} {r_ + }^4}}{{512{\mkern 1mu} \mu }} + \frac{{\pi {\mkern 1mu} {r_ + }^3}}{{1024{\mkern 1mu} \mu }} + \frac{{3{\mkern 1mu} \pi {\mkern 1mu} {r_ + }^2}}{{16384{\mkern 1mu} \mu }} + \frac{{3{\mkern 1mu} \pi {\mkern 1mu} {r_ + }}}{{65536{\mkern 1mu} \mu }}\\
+ \left( {3/8{\mkern 1mu} \frac{{\pi {\mkern 1mu} {r_ + }^4}}{\mu } + 1/16{\mkern 1mu} \frac{{\pi {\mkern 1mu} {r_ + }^3}}{\mu } + \frac{{27{\mkern 1mu} \pi {\mkern 1mu} {r_ + }^2}}{4} + \frac{{3{\mkern 1mu} \pi {\mkern 1mu} {r_ + }^2}}{{256{\mkern 1mu} \mu }} + \frac{{9{\mkern 1mu} \pi {\mkern 1mu} {r_ + }}}{8} + \frac{{3{\mkern 1mu} \pi {\mkern 1mu} {r_ + }}}{{1024{\mkern 1mu} \mu }}} \right)\alpha  + O\left( {{\alpha ^2}} \right)
\end{array}
\end{equation}
where we neglect the terms related to ${{\alpha ^2}}$ as it is so small. In this case, it is interesting that increasing of the $\alpha$ caused increase of the entropy (see figure 8).
\begin{figure}[tpb]
	\centering \begin{center} \end{center} 
	\includegraphics[width=1\textwidth,trim=-200 400 0 100,clip]{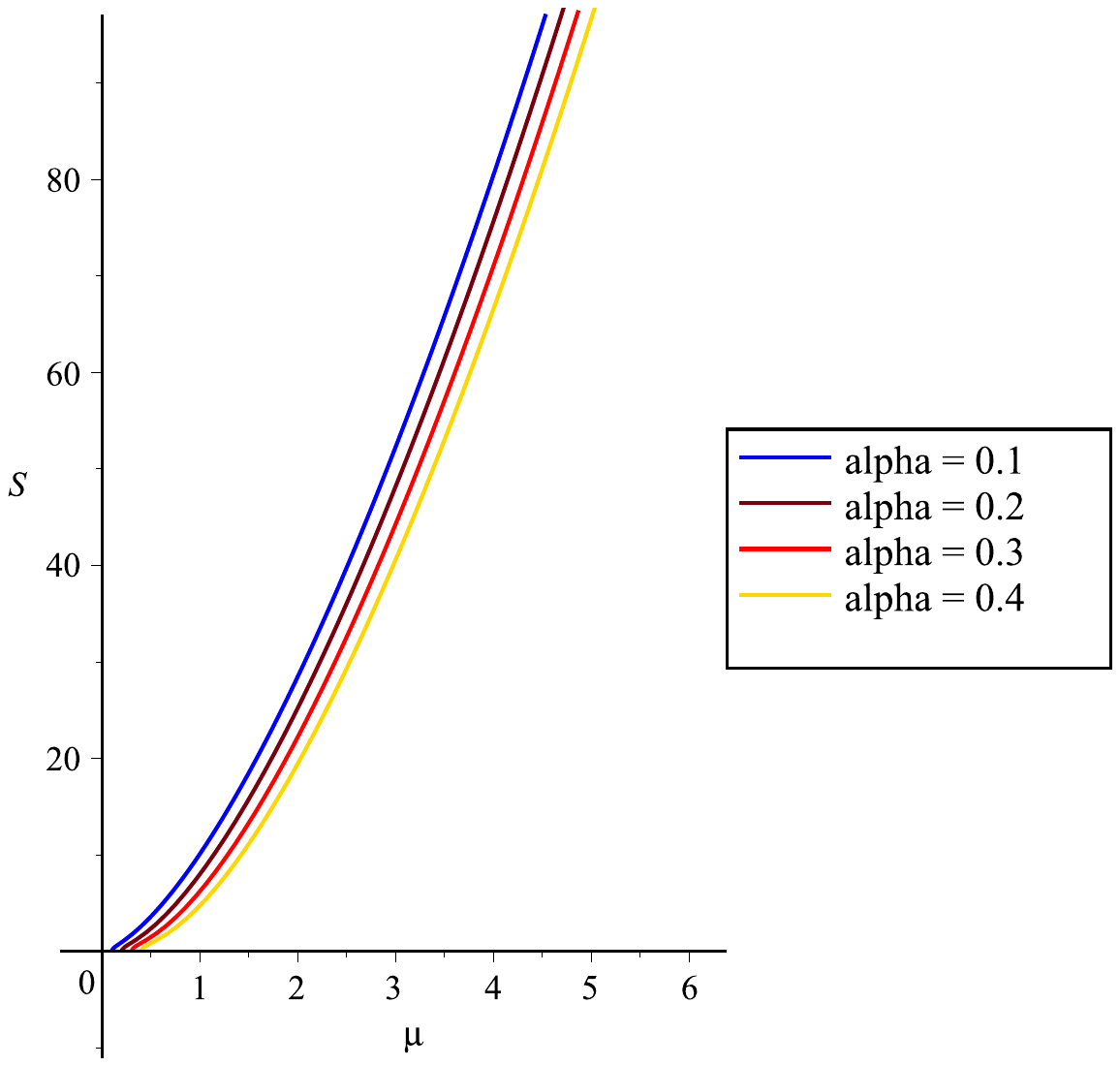}
	\hfill
	\caption{\label{fig6} The effect of $\alpha$ on the Bekenstein-Hawking entropy for $n = 3$}
\end{figure}
\begin{figure}[tpb]
	\centering \begin{center} \end{center} 
	\includegraphics[width=1\textwidth,trim=-200 400 0 100,clip]{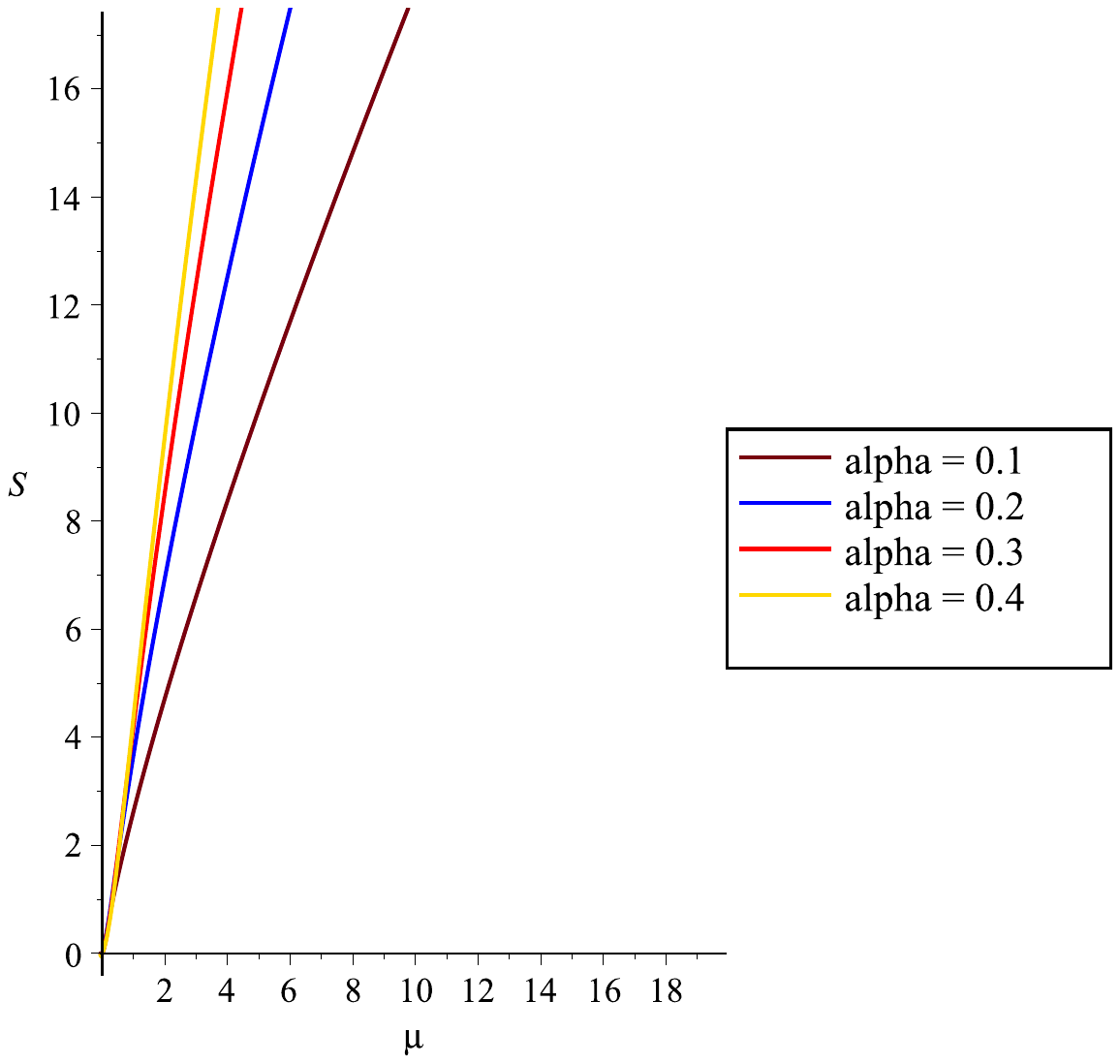}
	\hfill
	\caption{\label{fig6} The effect of $\alpha$ on the Bekenstein-Hawking entropy for $n = 4$}
\end{figure}  
If we set $n = 5$ by using equations (15) and (16), the entropy of the 7 dimensional spacetime, ${S^7}$, reads
\begin{equation} \label{3.12}
{S^7} = \int\limits_0^{{r_ + }} {{\mkern 1mu} \left[ {\frac{{648\pi {\mkern 1mu} \left( {{r^2} + 3{\mkern 1mu} \alpha } \right)\left( {\alpha {\mkern 1mu} {r^6} + 4{\mkern 1mu} \beta {\mkern 1mu} \mu } \right){{\left( {\beta {\mkern 1mu} \left( {\alpha {\mkern 1mu} {r^6} + 4{\mkern 1mu} \beta {\mkern 1mu} \mu } \right)} \right)}^{4/3}}}}{{81{\mkern 1mu} {{\left( {\beta {\mkern 1mu} \left( {\alpha {\mkern 1mu} {r^6} + 4{\mkern 1mu} \beta {\mkern 1mu} \mu } \right)} \right)}^{4/3}}\alpha {\mkern 1mu} \mu  - \sqrt[3]{{54}}{{\left( {\alpha {\mkern 1mu} \beta {\mkern 1mu} {r^6}} \right)}^{2/3}}\beta {\mkern 1mu} {r^{10}} - 27{\mkern 1mu} \sqrt[3]{{54}}{{\left( {\alpha {\mkern 1mu} \beta {\mkern 1mu} {r^6}} \right)}^{2/3}}\alpha {\mkern 1mu} \beta {\mkern 1mu} \mu {\mkern 1mu} {r^4}}} + O\left( {{\alpha ^{n \succ 3}},{\beta ^{n \succ 3}}} \right)} \right]{\mkern 1mu} {\rm{d}}r}  
\end{equation}
Solving (21) analytically is not straightforward but one can use some numerical methods in order to compute the entropy. However, we can neglect $\alpha$ and $\beta$ higher than order of 3. Obviously, the entropy for 7 dimensional asymptotically flat black hole in Lovelock back ground is related to the second and third order of the coupling constant of the Lovelock gravity theory. 
\section{Conclusion}
In this paper, we have calculated the Hawking temperature (equations (11), (12), (15)) and the Bekenestein-Hawking entropy (equations (19), (20), (21)) of the asymptotically flat black hole in Lovelock background for 5,6, and 7 dimensional spacetime by using a general algebraic equation (9). In this case, the coupling constant of the Lovelock gravity theory take into account up to the third order. It is evident that the thermodynamic parameters, like Hawking temperature and the entropy, are related to the second order of the coupling constant of the Lovelock for $n = 3,4$. In this manner, for 7 dimensional case, they are related to the second and third order of the constant of the Lovelock.
\\We have found, in case of $n = 3$ the maximum of the temperature decreases when the $\alpha$ increase as well as the entropy in a fixed mass. For 6 dimensional case, the temperature decreases with increasing $\alpha$. We have investigated the effects of the coupling constants of the Lovelock theory on the temperature and entropy in $d = 7$ as well. We have calculated the temperature analytically but we found that the entropy equation (21) needs to be solved numerically. In this case, both the temperature and the entropy are related to the second and third order of the coupling constant of the Lovelock gravity theory. It is evident the Hawking temperature increases when the $\alpha$ increases but it decreases with increasing $\beta$ (see figure 4 and 5). We have shown that variation of the $\alpha$ is more effective on the thermodynamics of the black holes rather than $\beta$.  

\section*{Acknowledgment}

The paper is supported by Universiti Kebangsaan Malaysia (Grant No. FRGS/2/2013/ST02/UKM/02/2) and University of Malaya (Grant No. Ru-023-2014).


\end{document}